\documentclass{emulateapj}

\def\etal{et~al.\ }

\slugcomment{Version: Feb 9, 2006}

\shorttitle{Red Galaxies at $z \sim 0.9$}
\shortauthors{Schiavon et al.}

\begin{document}


\title{The DEEP2 Galaxy Redshift Survey: Mean Ages and Metallicities of
Red Field Galaxies at $z \sim 0.9$ from Stacked Keck/DEIMOS Spectra$^1$}


\author{Ricardo P. Schiavon\altaffilmark{2}, 
S. M. Faber\altaffilmark{3},
Nicholas Konidaris\altaffilmark{3}, 
Genevieve Graves\altaffilmark{3},
Christopher N.A. Willmer\altaffilmark{3,4}, 
Benjamin J. Weiner\altaffilmark{5},
Alison L. Coil\altaffilmark{4,6,8}, 
Michael C. Cooper\altaffilmark{6},
Marc Davis\altaffilmark{6}, 
Justin Harker\altaffilmark{3}, 
David C. Koo\altaffilmark{3}, 
Jeffrey A. Newman\altaffilmark{6,7,8}
\& Renbin Yan\altaffilmark{6}}
\affil{$^2$ Department of Astronomy, University of Virginia, P.O. Box
3818, Charlottesville, VA 22903-0818}
\affil{$^3$ UCO/Lick Observatory/Department of Astronomy and Astrophysics, 
University of California, Santa Cruz, CA 95064}
\affil{$^4$ Steward Observatory, University of Arizona, Tucson, AZ 85721}
\affil{$^5$ Department of Astronomy, University of Maryland, College Park,
MD 20742-2421}
\affil{$^6$ Department of Astronomy, University of California, Berkeley,
601 Campbell Hall, Berkeley, CA 94720-3411}
\affil{$^7$ Lawrence Berkeley National Laboratory, Berkeley, CA 94720}
\affil{$^8$ Hubble Fellow}



\altaffiltext{1}{Based on observations taken at the W. M. Keck Observatory}


\begin{abstract}

As part of the DEEP2 galaxy redshift survey, we analyze absorption line
strengths in stacked Keck/DEIMOS spectra of red field galaxies with
weak to no emission lines, at redshifts 0.7 $\leq z \leq$ 1. Comparison
with models of stellar population synthesis shows that red galaxies at
$z \sim 0.9$ have mean luminosity-weighted ages of the order of only
1 Gyr and at least solar metallicities. This result cannot be reconciled
with a scenario where all stars evolved passively after forming at {\it very}
high $z$.  Rather, a significant fraction of stars can be no more than 1
Gyr old, which means that star formation continued to at least $z \sim
1.2$. Furthermore, a comparison of these distant galaxies with a local
SDSS sample, using stellar populations synthesis models, shows that the
drop in the equivalent width of $H\delta$ from $z \sim 0.9$ to 0.1 is
less than predicted by passively evolving models. This admits of two
interpretations: either each individual galaxy experiences continuing
low-level star formation, or the red-sequence galaxy population from
$z \sim 0.9$ to 0.1 is continually being added to by new galaxies with
younger stars.

\end{abstract}





\section{Introduction}

The formation of early-type galaxies is one of the ongoing riddles of
modern extragalactic astrophysics. According to the leading models,
massive early-type galaxies have been assembled hierarchically, from the
merging of less massive structures. Because such mergers are seen locally
to be accompanied by star formation (e.g., Schweizer \& Seitzer 1992),
one of the best ways to test the hierarchical formation paradigm
is by determining the star formation history of early-type galaxies.
This can be achieved by estimating the ages of stars in galaxies from
their integrated light, through comparison with stellar population
synthesis models.  Several groups have attempted to achieve this goal
from observations of massive galaxies in a range of redshifts (e.g.,
Le Borgne \etal 2006, Treu \etal 2005, Daddi \etal 2005, Longhetti \etal
2005, and references therein).  However, spectroscopic dating of stellar
populations older than $\sim$ 1 Gyr is best achieved by simultaneously
matching the strengths of Balmer and metal lines in their integrated
spectra, in order to avoid spurious effects due to the
age-metallicity degeneracy. So far, observational difficulties have
prevented such detailed studies for all but local samples (e.g., Gonz\'alez
1993, Trager \etal 2000, Kuntschner 2000, Caldwell \etal 2003, Eisenstein
\etal 2003, Thomas \etal 2005, Schiavon 2006, and references therein).

In this {\it Letter} we present the analysis of absorption line
strengths measured in stacked integrated Keck/DEIMOS spectra of red
galaxies with redshifts between 0.7 and 1, as part of the DEEP2 survey
(Davis \etal 2003).  We find that the stars in these galaxies have mean
light-weighted single stellar population (SSP) ages of order only 1 Gyr,
and their metallicities are at least solar. Since these objects are
observed several billion years after the big bang, this result suggests
that stars inhabiting red galaxies were formed during an extended
period of time.


\section{Sample and Data} \label{data}

The data used in this {\it Letter} consist of k-corrected absolute
magnitudes in the Vega system and 1-hour exposure Keck/DEIMOS (Faber
\etal 2003) spectra from DEEP2 (Davis \etal 2003). Redshift
determinations are described in Davis \etal (2003), and restframe $M_B$
magnitudes and U--B colors were derived from CFHT {\it BRI} photometry
and redshifts by Willmer \etal (2006). The S/N of each 1-hour exposure
spectrum is not high enough for accurate measurement of absorption line
indices, so we stack spectra of hundreds of galaxies, selected in bins
of color, luminosity, and redshift. Further details can be found in
Schiavon \etal (in preparation).

\subsection{Sample Selection}

\begin{figure}
\epsscale{1.2}
\plotone{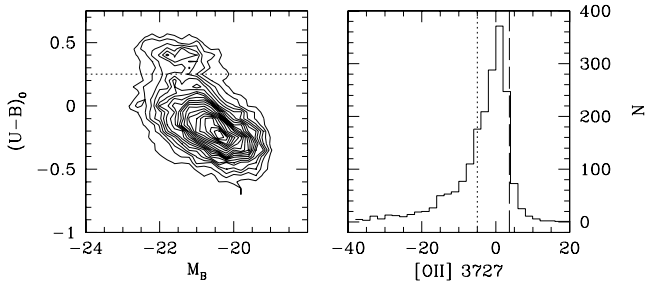}
\caption{{\it Left panel:} Restframe color-magnitude diagram of 17,745
DEEP2 galaxies. Our sample of RSGs is defined by U--B $>$ 0.25 (objects
above the dotted line). {\it Right panel:} A histogram of [OII] $\lambda$
3727 EWs for RSGs.  No [OII] emission happens when EW[OII]=3.7${\rm\AA}$
(dashed line) and lower values indicate the presence of line emission. All
galaxies with EW[OII] $\leq$ --5 ${\rm\AA}$ are excluded from the sample.
\label{fig1}} 
\end{figure}

Our goal is to study the evolution of red-sequence early-types, so we
first select galaxies by color. This selection criterion is illustrated
in the left panel of Figure~\ref{fig1}, where data for 17,745 DEEP2
galaxies with $0.7 \leq z \leq 1.05$ were used to produce a contour
plot on the restframe color-magnitude space. Red-sequence galaxies
(RSGs) are chosen to be those with U--B $\geq$ 0.25, making up a total
of 1941 objects. Ideally, we would also like to select galaxies on the
basis of morphology, but we unfortunately lack that information for the
sample under analysis. Therefore, in order to minimize contamination
by strongly reddened late-type galaxies we impose another cut, based on
the equivalent width (EW) of [OII] 3727 (see definition by Fisher \etal
1998). This is illustrated in the right panel of Figure~\ref{fig1},
where a histogram of [OII] EWs is shown for all RSGs in our sample.
Strong emission-line RSGs have very negative values of EW[OII] while
the zero of [OII] emission is at $\sim$ 3.7 ${\rm\AA}$ (Konidaris \etal
in preparation). The distribution is strongly peaked at very low [OII]
emission values with a long tail towards galaxies with strong [OII]
emission.  Contamination by reddened late-type galaxies is likely to be
more important in the strong-emission line regime, so that we remove
from our sample all galaxies with [OII] EW $\leq$ --5 ${\rm\AA}$.
This emission-line cut admittedly leaves in our sample a large number
of galaxies in the low-emission line regime. These are mostly AGN on
the basis of the ratios between [OII] and residual Balmer line emission
(Schiavon \etal in preparation, Konidaris \etal in preparation), as has
been also found for low redshift RSGs (Phillips \etal 1986, Rampazzo
\etal 2005, and Yan \etal 2006).

The color and emission-line cuts leave us with a sample of 1160 galaxies.
We create six subsamples out of this set of galaxies, three with varying
colors but the same redshift range, and three with varying redshifts,
but consistent colors and luminosities.  The color and redshift limits of
each bin are listed in Table~\ref{tbl1}. In an attempt to compare objects
with similar masses, galaxies in the color and redshift sub-samples
were further selected within 1 mag-wide $M_B$ intervals where the
central magnitude was chosen to be consistent with passive evolution
from the age and metallicity of the high-$z$ sample. However,
adopting the exact same $M_B$ interval for each $z$ bin does not
change the results. The numbers of galaxies in each bin are listed in
Table~\ref{tbl1}.

\begin{deluxetable*}{lccccccc}
\tablecaption{Data for galaxies used in stacked DEEP2 spectra\label{tbl1}}
\tablewidth{0pt}
\tablehead{
\colhead{Bin} & \colhead{$z$} & \colhead{$M_B$/$M_r$} & \colhead{$U-B$} & 
\colhead{$H\delta_F$} & \colhead{Fe4383} & \colhead{$\sigma$ (km/s)} & \colhead{N} 
}
\startdata
Low z & [0.7,0.8] & [--21.57,--20.57] & [0.25,0.60] & 1.8$\pm$0.1 & 3.7$\pm$0.5 & 190 & 113\\
Intermediate z & [0.8,0.9] & [--21.70,--20.70] & [0.25,0.60] & 1.8$\pm$0.1 & 3.0$\pm$0.2 & 170& 288\\
High z & [0.9,1.0] & [--21.86,--20.86] & [0.25,0.60] & 1.8$\pm$0.2 & --- & 180 & 167\\
Red & [0.75,0.95] & [--21.76,--20.76] & [0.45,0.60] & 1.6$\pm$0.2 & 3.7$\pm$0.3 & 190 &
129\\
Intermediate & [0.75,0.95] & [--21.76,--20.76] & [0.35,0.45] & 1.8$\pm$0.1 & 3.2$\pm$0.2 & 170 &
228\\
Blue & [0.75,0.95] & [--21.76,--20.76] & [0.25,0.35] & 2.3$\pm$0.2 & 2.4$\pm$0.3 & 170 &
119\\
SDSS - Lum~215 & 0.171 & [--22.0,--21.5] & -- & 0.66$\pm$ 0.01 & 4.32$\pm$0.02 & 235 & 5412 \\
SDSS - Lum~210 & 0.143 & [--21.5,--21.0] & -- & 0.82$\pm$ 0.01 & 4.22$\pm$0.02 & 210 & 6477 \\
\enddata
\tablecomments{Numbers in brackets correspond to intervals adopted
to select galaxies in different bins. Single numbers correspond to
measurements taken in stacked spectra or, in the case of U--B for DEEP2
and $z$ for SDSS, average values within a given bin. $N$ is the number
of galaxies in a bin. Absolute magnitudes are $M_B$ for DEEP2 and $M_r$
for SDSS.
} \end{deluxetable*}
\subsection{Stacked Spectra and Lick Indices}

The 1160 galaxy spectra were visually inspected in order to clean the
sample from a few misclassified stars, galaxies with wrong redshifts,
and zero-S/N spectra. A rough relative fluxing was achieved by dividing
each spectrum by the normalized throughput of the DEIMOS spectrograph
with the 1200 l/mm grating.  Before coaddition, the spectra were
brought to restframe and then normalized through division by the
average ($\sigma$-clipped) counts within the $\lambda\lambda$ 3900--4100
${\rm\AA}$ interval. Coaddition was performed adopting a $\sigma$-clipping
procedure to eliminate sky-subtraction residuals, zero-count pixels
due to CCD gaps, and other spectral blemishes. After several tests the
best results were obtained when a single 3-$\sigma$ clipping iteration
was adopted. On average more than 90\% of all galaxies in a given bin
contribute to the stacked spectrum at any given wavelength. No clipping
was performed in the region of the [OII] $\lambda$ 3727 ${\rm\AA}$
line. In Figure~\ref{fig2}a we compare one of our stacked spectra with
a SSP model from Schiavon (2006). In order to match the overall flux
distribution of the theoretical spectrum, the observed spectrum was
dereddened by E(B--V)=0.2. Since the observations were not properly
flux-calibrated, this E(B-V) value does not reflect the average 
reddening in the sample galaxies and this correction has the sole
purpose of bringing observations and theory to a common relative scale
so as to highlight the outstanding agreement between line strengths
in the observed and synthetic spectra.

All Lick indices in the $\lambda\lambda$ 4000--4500 ${\rm\AA}$
region were measured in the stacked spectra, but we focus here on the
$H\delta_F$ and Fe4383 indices, which are chiefly sensitive to age
and [Fe/H], respectively. The spectra were first broadened to the Lick
resolution as given by Worthey \& Ottaviani (1997), and the indices were
measured following definitions by those authors and by Worthey \etal
(1994).  Velocity dispersions ($\sigma$) were measured in the stacked
spectra through Fourier cross correlation using the IRAF {\tt rv.fxcor}
routine. The template adopted was a model spectrum from Schiavon (2006)
for a SSP with solar metallicity and an age of 2 Gyr. The same model
spectrum was used to infer corrections to the line indices for the effect
of $\sigma$-broadening. The indices were all corrected to $\sigma =0$
km/s using the $\sigma$ determined for each stacked spectrum. The latter
are listed in Table~\ref{tbl1}. We do not attempt to convert the line
indices to the Lick system, aside from smoothing them to the Lick/IDS
resolution. However, zero-point differences should be very small, given
that the Schiavon (2006) models are based on fluxed spectra and the DEEP2
spectra are corrected from instrumental throughput. Finally, Balmer
lines were corrected for emission-line in-fill, which was estimated
from EW[OII], adopting EW[OII]/EW($H\alpha$) = 6 (Yan \etal 2006)
and EW($H\delta$) = 0.13 EW($H\alpha$).  The correction to $H\delta_F$
is smaller than 0.2 ${\rm\AA}$, corresponding to less than 1 Gyr in age.

\subsection{Local Sample}

Galaxy evolution is better assessed when distant and local samples of
similar objects are contrasted using evolutionary models.  Moreover, it
is vital that the nearby and distant samples are defined as consistently
as possible, to ensure that the two samples represent objects of the same
class. For a local counterpart to the distant DEEP2 sample we use the SDSS
data from Eisenstein \etal (2003), who provide stacked flux-calibrated
spectra of RSGs, binned by absolute magnitude, environment, and
redshift. Because the DEEP2 stacked spectra include galaxies from all
environments, we chose to use stacked spectra from a similarly defined
sample from Eisenstein \etal (their ``All'' sample). Furthermore, in
order to match the relative position of our absolute magnitude bins
along the red sequence, we chose to exclude both the lowest and highest
magnitude bins of the Eisenstein \etal sample from our analysis. The
spectra were downloaded from D. Eisenstein's website and submitted to
the same treatment as described above for the DEEP2 spectra. Key data
for the Eisenstein \etal (2003) sample are listed in Table~\ref{tbl1}

\section{Mean Ages and Metallicities} \label{agemet}

\begin{figure*}
\plotone{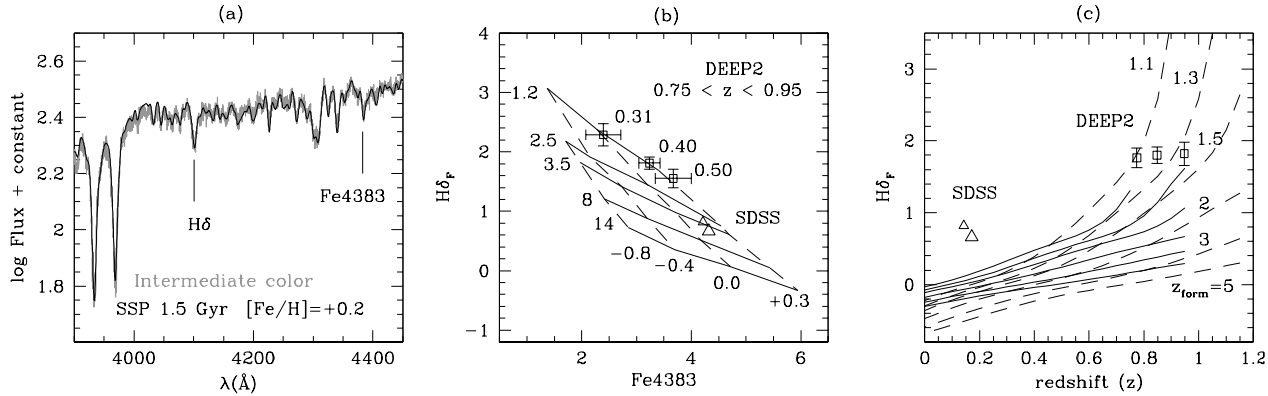}
\caption{a) Comparison between the stacked spectrum for the intermediate
color bin (see Table~\ref{tbl1}) and that for a model SSP from Schiavon
(2006). The two absorption lines studied here are indicated. This plot
illustrates the high S/N of the stacked spectra and also the high
quality of the models, which reproduce all the main absorption lines
in the observed spectrum very well. b) Measurements from RSG
stacked spectra versus predictions of SSP models.  Squares with error
bars indicate DEEP2 galaxies and triangles indicate SDSS galaxies
from Eisenstein et al. (2003). For SDSS data, symbol size correlates
with luminosity. DEEP2 galaxies are binned by color and have absolute
magnitude and redshifts within the intervals indicated in the label
and in Table~\ref{tbl1}. The mean colors of DEEP2 bins are indicated
to the right of each data point. The models are those of Schiavon
(2006) for [$\alpha$/Fe] = +0.4. Same-age (same-metallicity) models are
connected by solid (dashed) lines. Model age ([Fe/H]) ranges from 1.2
to 14 Gyr (-0.8 to +0.3 dex), as indicated by the labels.  The mean SSP
ages of DEEP2 galaxies are $\sim$ 1.2 Gyr, regardless of color. They
are younger than the SDSS galaxies, as expected. Their [Fe/H]s range
between solar and +0.3. 
c) DEEP2 data from the redshift-selected sample compared with SSP
models in a redshift vs.\ $H\delta_F$ plot. Predictions from SSP
models with super-solar metallicity and a range of formation redshifs,
$z_{form}$, are shown as solid ([Fe/H]=+0.2, [$\alpha$/Fe]=0) and dashed
([Fe/H]=+0.3, [$\alpha$/Fe]=+0.4) lines.  DEEP2 galaxies are consistent
with $z_{form}$ ranging from 1.1 to 1.3 (1.3 to 1.5 if [Fe/H]=0 SSP models
are adopted). This indicates that star formation was probably prolonged in
those galaxies. Galaxies at high and low $z$ are {\it not} connected by
lines of passive evolution, which probably indicates that star formation
did not cease from $z \sim 0.9$ to the present day.  \label{fig2}}
\end{figure*}

Lick indices measured in the stacked spectra are compared with SSP models
in Figure~\ref{fig2}b. Shown are the indices measured in Eisenstein \etal
(2003) spectra and those from the color-selected DEEP2 sub-samples. Symbol
size for the Eisenstein \etal galaxies is proportional to luminosity. The
data are compared to models computed adopting [$\alpha$/Fe] = +0.4,
[C/Fe]=+0.15, and [N/Fe]=+0.3 (See Schiavon \etal in preparation for
details).

The main result of this {\it Letter} is immediately apparent in
Figure~\ref{fig2}b. The stellar populations of field RSGs at $z \sim$ 0.9
are young, with a mean luminosity-weighted age of only $\sim$ 1.2 Gyr.
Adoption of solar-scaled models would imply older ages by roughly
0.5 Gyr. Iron abundances range from [Fe/H] $\sim$ 0 to $\sim$ +0.3.
As expected, the stellar populations in the local SDSS galaxies are older
than those at $z \sim 0.9$, with mean ages ranging between 3 and 6 Gyr,
with above solar [Fe/H]. The time difference from the characteristic
redshifts of DEEP2 to SDSS samples is over 5 Gyr, so the expected mean
ages of SDSS galaxies under passive evolution should be over 6 Gyr.
We note that the Schiavon (2006) models match the same data for Galactic
globular clusters and---most importantly for this study---those for the
open cluster M~67 (3.5 Gyr, [Fe/H]=0), to within 1 Gyr in age
and 0.1 dex in [Fe/H].

Another interesting way of viewing this result is illustrated in
Figure~\ref{fig2}c, where DEEP2 and SDSS galaxies are compared with
passively evolving SSP models in a redshift vs.\ $H\delta_F$ plot. The
DEEP2 data plotted in this case come from the redshift-selected sub-sample
(see Table~\ref{tbl1}). The lines correspond to model predictions for
SSP evolution assuming various redshifts of formation ($z_{form}$) and
adopting a concordance WMAP cosmology (Spergel \etal 2003). Dashed (solid)
lines represent models with [$\alpha$/Fe] = 0.0 (+0.4) and super-solar
[Fe/H]. Figure~\ref{fig2}c shows that the data for field RSGs require
$z_{form} \sim$ 1.1--1.3, when they are
modeled using SSPs with super-solar metallicity. Moreover, the distant
and local samples are not connected by lines of passive evolution, which
indicates the occurrence of star formation between $z \sim$ 0.9 to $\sim$
0.1 (see also Gebhardt \etal 2003). This could be due either to {\it in
situ} star formation or to the incorporation in the red sequence of new
galaxies coming from the blue cloud after cessation of star formation.
Figure~\ref{fig2}c also makes clear that these results are valid
regardless of the [$\alpha$/Fe] ratio of the models adopted. The results
also remain qualitatively unchanged when models with lower metallicity
are used. Adoption of solar metallicity models (not shown) would result
in slightly higher redshifts of star formation, $z_{form} \sim$ 1.3--1.5.

\section{Conclusions and Caveats} \label{conc}

The results presented in this {\it Letter} need to be interpreted with
caution.  Ages and metallicities inferred from comparison of integrated
galaxy spectra with SSP models are luminosity-weighted averages, whereas
the real stellar populations doubtless consist of stars with a range of
ages. Therefore, we are not claiming that the stars in the DEEP2 galaxies
plotted in Figure~\ref{fig2}b are uniformly $\sim$ 1.2 Gyr old. Likewise,
we are not proposing that the mean galaxies plotted in Figure~\ref{fig2}c
sprang into existence at $z \sim 1.3$. But the strength of $H\delta$
in the stacked integrated spectra of field RSGs indicates that they
harbour young and/or intermediate-age stars both locally and at $z
\sim 0.9$. Since the universe was roughly 6 Gyr old at $z \sim 0.9$,
this result cannot be reconciled with models in which all the stars in
these galaxies were formed at very high redshifts ($z > 3$) and evolved
passively ever since. In fact, it appears that star formation in these
galaxies was prolonged, and that it continued, in small amounts,
between $z \sim 0.9$ and 0.1. The fact that we see no evolution in 
[Fe/H] seems to indicate that the bulk of star formation has occurred
before $z \sim 0.9$.

The presence of young/intermediate-age stars in early-type galaxies can
be accounted for by at least two scenarios. The so-called {\it frosting}
models (e.g., Trager \etal 2000) propose small amounts of recent {\it
in situ} star formation originated these stars. On the other hand, {\it
quenching} models (e.g., Bell \etal 2004, Faber \etal 2005) suggest
that blue galaxies migrate to the red sequence after cessation of star
formation, possibly associated with a merger event and/or enhanced AGN
activity.  In a separate study (Harker \etal in preparation) we make an
exploration in that direction, by showing that quenching models provide
a good match to the data analyzed in this {\it Letter}. Quenching models
were also shown by Faber \etal (2005) to provide a good match to the
evolution of the luminosity function of red galaxies from $z \sim 1.4$
to the present day.  An estimate of the fraction of the total stellar
mass allocated in these young/intermediate-age stars (e.g., Leonardi \&
Rose 1996) might help discriminate between the different scenarios.

A few important caveats must be kept in mind when considering these
results. Of most importance, our sample is not yet selected on the basis
of morphology. Although contamination by late-type, star-forming galaxies
is low (Konidaris \etal 2005 in preparation), a stronger statement on
the history of star formation of early-type galaxies awaits an analysis
of a morphologically selected sample of distant objects (e.g., Treu
\etal 2005). It is also important to keep in mind that we are dealing
with a field sample. Previous studies of cluster samples at comparable
redshifts tell a different story, with cluster RSGs being compatible
with high $z_{form}$ (e.g., Kelson \etal 2001).
Lastly, the local SDSS sample used here does not match perfectly the
selection criteria adopted for our DEEP2 sample. Work aimed at producing
more adequate local counterparts to our distant DEEP2 sample is currently
under way (Graves \etal 2005 in preparation).

\acknowledgments

This project was supported in part by NSF grants AST 00-71198 and AST
00-71408, and AST-0507483. R.P.S. acknowledges financial support from
HST Treasury Program grant GO-09455.05-A to the University of Virginia.
R.P.S. thanks Bob O'Connell and Jim Rose for helpful comments on an early
version of the manuscript.  J.A.N. acknowledges support from NASA through
Hubble Fellowship grant HST-HF-01165.01-A awarded by the Space Telescope
Science Institute, which is operated by the Association of Universities
for Research in Astronomy, Inc., for NASA, under contract NAS 5-26555.
We thank the Hawaiian people for allowing us to conduct observations
from their sacred mountain.

{}








\end{document}